\documentstyle[12pt]{article}
\newcommand{\beq}{\begin{equation}}
\newcommand{\eeq}{\end{equation}}
\newcommand{\bdis}{\begin{displaymath}}
\newcommand{\edis}{\end{displaymath}}
\newcommand{\bea}{\begin{eqnarray}}
\newcommand{\eea}{\end{eqnarray}}
\newcommand{\barr}{\begin{array}}
\newcommand{\earr}{\end{array}}
\newcommand{\beas}{\begin{eqnarray*}}
\newcommand{\eeas}{\end{eqnarray*}}

\begin{document}
\begin{center}
{\large{\bf New connections between moving curves and soliton
 equations }}

\vskip .4cm

{\bf S. Murugesh,} ~~
${\bf Radha Balakrishnan^*}$

\vskip .3cm

{\it The Institute of Mathematical Sciences, C.I.T. Campus, Chennai 600
113, India}

\end{center}

\bigskip
\hrule
\bigskip
\baselineskip=20pt

\noindent {\bf Abstract}\\

  Lamb  has identified a certain  class of moving space  curves
with soliton equations. We show that there are  two other classes of curve
evolution  that may be so identified. Hence
  {\it three} distinct classes of curve evolution  are  associated
 with a given  integrable equation.
 The  nonlinear Schr\"{o}dinger equation is used to illustrate this
   explicitly.\\

\vskip 1.5 cm
\hrule
\smallskip
\noindent $^{*}~ {\small\rm Corresponding~~ author.~~Tel:
91-44-2351856~~;~~fax: 91-44-2350586}$ \\
\noindent{\it E-mail~~ address}:~~{\small \rm radha@imsc.ernet.in}\\
\newpage
\section{Introduction}

Over two decades ago, Lamb \cite{lamb} presented a formalism
 which  showed that certain special types of motion of
 space curves  can be mapped to completely integrable,
 soliton-supporting  \cite{ablo}, nonlinear partial differential equations
(NLPDE) such as the nonlinear Schr\"{o}dinger (NLS) equation,
 the sine-Gordon
equation, the Hirota equation,
etc., indicating that the corresponding curve motions are also
integrable. This formalism arose as an extension of
  Hasimoto's  earlier work \cite{hasi}, which had established a
 connection between the equation of motion of a vortex filament
regarded as a moving space curve, and the NLS. Since all
 soliton equations \cite{ablo} possess similar
characteristics such a Lax pair, solvability by the
inverse scattering transform method and an infinite number of constants of
motion, the above mapping to the NLS  provided a motivation for
  a general formulation to investigate the interesting  curve
 evolutions  that get associated with  soliton-supporting
 equations.

Central to the Lamb formulation is the introduction of the
following  complex function $\psi$, called the Hasimoto
function, defined \cite{hasi} as
\beq
\psi(s,u) \ = \ K  \exp { [i\int \tau ds] }.
\eeq

Here  $s$ is the arc length of the  curve and $u$ the time.
 $K(s,u)$
and $\tau(s,u)$ denote, respectively, the curvature and torsion of
 the moving curve.
 It is
 this function $\psi$ that satisfies
  various  integrable equations in Lamb's work, and is
 used  frequently in the study of various aspects of
curve-dynamics \cite{klapper}.

   It is clear  that by comparing the functional form
of $\psi$ (Eq. (1.1) with a soliton solution
 of a given integrable equation for $\psi$,  one
 can  identify $K$ and $\tau$, and hence obtain the associated
moving curve parameters that correspond to  an integrable,
shape-preserving  curve motion. This in turn unravels  a certain
   special  geometric structure of the given
soliton-bearing NLPDE. In this Letter, we show that two other
 complex functions defined by
$\Phi(s,u) \ = \ \tau  \exp {[i\int K  ds]}$
and $\chi \ = \ (K - i \tau)$,
 also arise from the basic curve evolution equations, just as
naturally as the
Hasimoto function $\psi$ does.
 We demonstrate that
these  can also satisfy  various  soliton equations.
This in turn leads to the result  that each such integrable
  equation  is associated with not just one, but
{\it three} distinct classes of space curve motion, and therefore
 has a much {\it richer} geometric structure than
hitherto envisaged.
We illustrate this by using the NLS as an example.

\section{Moving space curves}
\setcounter{equation}{0}
 A space curve embedded in three-dimensions may be
described using the usual Frenet-Serret equations \cite{eisen}
 for the orthonormal triad of unit vectors   made up of
the tangent {\bf t}, normal {\bf n} and the binormal {\bf b} :
\beq
{\bf t}_s \ = \ K {\bf n} \ ; \ {\bf n}_s  \ =
 \ -K{\bf t} \ + \ \tau
{\bf b}\ ; \ {\bf b}_s \ =\ -\tau {\bf n}.
\eeq
Here, $s$ stands for the arc length of the curve.The
  subscript  $s$  denotes $(\partial/\partial s)$.
 $K$ and $\tau$  are the curvature and torsion of the curve.

 If the curve moves with time
$u$, then all quantities in Eq. (2.1) become functions of
 {\it both} $s$ and $u$.
 To describe the time evolution of
 the triad $({\bf t}, {\bf n}, {\bf b})$
 we write down the following set of
equations  \cite{radha1}:
\beq
{\bf{t}}_u\,=\, g {\bf {n}} +h  {\bf{b}}; \quad  {\bf{n}}_u \,=\,
-  g {\bf{t}} + \tau_{0} {\bf{b}}; \quad  {\bf{b}}_u\,=\,-h
{\bf{t}} - \tau_{0} {\bf{n}}.
\eeq
Here, the subscript $u$ denotes $(\partial/\partial u)$.
 As is clear, the parametes $g,h$ and $\tau_0$  which determine
 the motion of the curve are also functions of both $s$
and $u$.

On requiring the compatibility conditions
\beq
{\bf t}_{su} \ = \ {\bf t}_{us} \ ;
 \ {\bf n}_{su} \ = \ {\bf n}_{us}
\quad ; \quad {\bf b}_{su} \ = \ {\bf b}_{us},
\eeq
a short calculation using Eqs. (2.1) and (2.2) leads to
\beq
K_u \ = \ (g_s - \tau h) \ ; \ \tau_u \ = \ (\tau_0)_s + Kh \
 ; \ h_s \ =
\ (K\tau_0 - \tau g).
\eeq

\section{Formulation (I) using the
Hasimoto function $\psi$.}
\setcounter{equation}{0}
In  Lamb's formulation \cite{lamb},
 which  will be referred to hereafter
 as formulation (I), the second and third
equations of the set (2.1) are combined to yield
\beq
({\bf n} + i{\bf b})_s \ + \ i \tau ({\bf n}
 + i{\bf b}) \ = \ - K{\bf t}.
\eeq
This immediately suggests the definition of a certain
 complex vector
\beq
{\bf N} \ = \ ({\bf n} + i{\bf b})
\exp {[i \int \tau ds]}.
\eeq
Differentiating Eq. (3.2) with respect to $s$ and using Eq.
(3.1), we get
\beq
{\bf N}_s \ = \ - K \exp [i \int \tau ds]{\bf t}.
\eeq
Thus the Hasimoto function  $\psi$  (Eq. (1.1))  appears in
 a natural fashion in the above equation.
Using the definitions of {\bf N} and $\psi$ given in Eqs. (3.2) and (1.1)
respectively, Eqs. (2.1)  can be written in the form
\beq
{\bf t}_s \ = \ \frac{1}{2} (\psi^* {\bf N} + \psi {\bf N}^*)
~~~;~~~{\bf N}_s \ = \ - \psi {\bf t}.
\eeq
Next, Eqs. (2.2) take on the form
\beq
{\bf t}_u \ = \ - \frac{1}{2} (\gamma^*_1 {\bf N} + \gamma_1 {\bf
N}^*)~~~;~~~
{\bf N}_u~~=~~ iR_1 {\bf N}~+~ \gamma_1 {\bf t},
\eeq
where
\beq
\gamma_{1} =-(g + ih) {\rm exp} [i \int \tau ds]~~~;~~~
R_1 \ = \ (\int \tau_u ds - \tau_0) \ = \ \int K h ds .
\eeq
Here we have used the second equation in the set (2.4) to write the
last equality in (3.6).  We use the  subscript $1$  here, to denote
 formulation (I). From Eqs. (3.4) and (3.5), setting ${\bf N}_{su} \ = \
{\bf N}_{us}$, and equating the coefficients of {\bf t} and {\bf N},
respectively, we get
\beq
\psi_u + \gamma_{1s} - i R_1 \psi \ = \ 0;
\eeq
\beq
R_{1s} \ = \ \frac{i}{2} (\gamma_1 \psi^* - \gamma_1^* \psi).
\eeq

 Interestingly,  as noted by Lamb \cite{lamb}, the structure of
Eqs. (3.7) and (3.8) which arose from
compatibility conditions on curve evolution,
 suggests  a possible relationship  with
soliton-bearing equations, via the Ablowitz-Kaup-Newell-Segur (AKNS)
 formalism \cite{ablo} . This is seen as
follows : It is well known that for a class of soliton-bearing
 equations with a function
$q(s,u)$ as the dependent variable, the Lax pair $L$ and $M$
 in the AKNS formalism are given by :
\beq
L~~y \ = \ \pmatrix{i \frac{\partial}{\partial s} & -iq \cr -iq* & -i
\frac{\partial}{\partial s}} y \ = \ \zeta~~y
\eeq
\beq
i~~ \frac{\partial y}{\partial u} \ = \ \pmatrix{A(s,u,\zeta) &
B(s,u,\zeta) \cr -B^{*}(s,u,\zeta) & -A(s,u,\zeta)}y \ = \ M~~y
\eeq
Here, the eigenfunction $y$ is the column vector $(y_{1}~~y_{2})^T$
 and $\zeta$ is the eigenvalue
parameter. Requiring $y_{su} = y_{us}$, Eqs. (3.9) and (3.10)  lead to the
following AKNS compatibility conditions \cite{ablo}:
\beq
q_u = 2 A q + B_s + 2i \zeta  B
\eeq
\beq
A_{s} = (B q^* - B^* q)
\eeq

Equations (3.11) and (3.12) are identical in form to Eqs. (3.7) and (3.8) provided  the
following identifications are made :
\beq
q = \psi /2;~~~A = i R_{1};~~~B= -\gamma_{1}/2;~~~\zeta=0.
\eeq
Thus the curve evolution equations (2.1) and (2.2) imply AKNS equations,
 with  $\zeta = 0$.
Now, $\gamma_1$ and $R_1$ are given in Eqs. (3.6).
 It was shown by Lamb \cite{lamb} that for appropriate choices of
 $\gamma_1$ as a function of $\psi$ and its derivatives, $R_{1}$
 can be found from Eq. (3.8), and substituted in
 Eq. (3.7)  to yield integrable equations for $\psi$.

 Alternatively, once  the quantities  $\gamma_{1}$ and $R_{1}$
are identified so that   Eqs. (3.7) and (3.8) take on the form of an
 integrable NLPDE for $\psi$, it is  possible to find the
 corresponding Lax pair, by
 working directly with equations (3.4)  for ${\bf t}_s$ and
${\bf N}_s$,  along with  equations (3.5)
for ${\bf t}_u$ and  ${\bf N}_u$ :  This is done \cite{lamb}  by
considering the constraint $t_{l}^2+n_{l}^2+b_{l}^2=1$, where
 the subscript
$l=1,2,3$ is used to represent the three components of the vectors
concerned. One then  defines a complex function
\beq
f^{(1)}_{l}= (n_{l}+ib_{l})/(1-t_{l})=(1+t_{l})/(n_{l}-ib_{l}),
\eeq
 along with its two other counterparts
 $f^{(2)}_{l}$ and $f^{(3)}_{l}$,
 obtained by cyclically changing $t$, $n$
and $b$ in  Eq. (3.14). It  can then be shown that  all the three
functions $f^{(\alpha)}_{l}$, $\alpha=1,2,3$ satisfy
 appropriate Riccati equations.  The  corresponding Lax pair in
each case can then be obtained by setting
$f^{(\alpha)}=y^{(\alpha)}_{2}/y^{(\alpha)}_{1}$,
 and identifying the
 corresponding eigenfunction to be the column vector
 $(y^{(\alpha)}_{1}~~y^{(\alpha)}_{2})^{T}$. A short calculation
shows that only one  of the Lax pairs thus obtained  has  the
 AKNS form of Eqs. (3.9) and (3.10),  with  entries as in (3.13).

 We paranthetically remark that it is possible
 to introduce a non-zero  eigenvalue parameter $\zeta$ in
 this formalism, by means of a
 suitable gauge transformation of $\psi$ in Eq. (3.7).
 This in turn  can be related  to a certain gauge freedom \cite{radha2}
 in the choice of the orientation of the  axes in the plane
 perpendicular to the tangent of the space
 curve. The details of this  and its ramifications  will
 be reported when completed.

\section{New formulations (II) and  (III) using
   functions  $\Phi$ and $\chi$}
\setcounter{equation}{0}

 As already noted, in Lamb's formulation
  or formulation (I),
 one proceeds by first combining the second and third equations of the
Frenet-Serret set of equations (2.1), leading to the appearance of
 $\psi$. In this section, we
consider the other two possibilities: Formulation (II), that  combines
  the first and second equations  of (2.1),
 and formulation (III) that  combines
 the first and third equations. As we shall
 see, these formulations
lead to the appearance of  two  other complex
functions $\Phi$ and $\chi$
respectively, in a natural fashion.

\noindent{\small {\bf Formulation (II):}} Combining the first two
 equations in Eqs. (2.1), we get
\beq
({\bf n} - i{\bf t})_s \ + \ i K ({\bf n} - i{\bf t}) \ = \ \tau {\bf
b}
\eeq
This suggests the definition of a {\it second} complex vector
\beq
{\bf M} \ = \ ({\bf n} - i{\bf t}) \ {\rm exp}  [i \int K ds]
\eeq
Differentiating Eq. (4.2) with respect to $s$ and using Eq. (4.1), we get
$ {\bf M}_s \ = \  \tau~~ {\rm exp} [i \int K ds] {\bf b}$.
Thus  a  {\it second} complex function
\beq
\Phi(s,u) \ = \ \tau  \exp { [i\int K ds] }
\eeq
  appears in a natural fashion in this case.
Using the definitions of {\bf M} and $\Phi$ given in Eqs. (4.2) and (4.3)
 in  the basic equations  (2.1) and (2.2),  and repeating the steps
 of Sec.3 , we obtain the following counterparts of Eqs. (3.4),(3.5) and
(3.6):
\beq
{\bf M}_s \ = \ \Phi {\bf b}~~~;~~~
{\bf b}_s \ = - \frac{1}{2} (\Phi^* {\bf M} + \Phi {\bf M}^*),
\eeq
\beq
{\bf M}_u = iR_2 {\bf M}  - \gamma_2 {\bf b}~~~;~~~
{\bf b}_u \ = \ \frac{1}{2} (\gamma^*_2 {\bf M} + \gamma_2 {\bf
M}^*)\,
\eeq
where
\beq
\gamma_{2} \ = - (\tau_{0} - ih) {\rm exp} [i \int K ds]~~~;~~~
R_2 \ =\ ( \int K_u ds - g ) \ = -\ \int \tau h ds.
\eeq
We have used the first  equation in the set (2.4) to write the
last equality in (4.6). The subscript $2$ is used on $\gamma$
and $R$ to indicate that these correspond to
 formulation (II). From Eqs.(4.4) and (4.5),
 setting ${\bf M}_{su} \ = \ {\bf M}_{us}$, and equating the
 coefficients of {\bf b} and {\bf M}, respectively, we get
\beq
\Phi_u + \gamma_{2s} - i R_2 \Phi \ = \ 0;
\eeq
\beq
R_{2s} \ = \ \frac{i}{2} (\gamma_2 \Phi^* - \gamma_2^* \Phi).
\eeq

 The  structure of Eqs. (4.7) and (4.8) is the  same  as
  that obtained in Lamb's formulation (see Eqs. (3.7) and (3.8)).  Thus using
 the same steps as in the  last section,  these can also be
  cast in the form of  AKNS compatibility conditions  (3.11) and (3.12),
 with the following {\it new} identifications:
\beq
q = \Phi /2;~~~A = i R_{2};~~~B= -\gamma_{2}/2;~~~\zeta=0.
\eeq
 This shows that, just as one does in Lamb's formulation,
 here also we could take  choices of $\gamma_2$, as  an
 appropriate function of $\Phi$ and its derivatives,
 find $R_{2}$ from (4.8), and substitute these expressions
 in Eq. (4.7) to obtain some of
 the well known integrable equations.

\noindent {\small {\bf Formulation (III):}} Combining the first and
 third equations of (2.1), we get
\beq
({\bf t} + i{\bf b})_s  \ = \ (K- i\tau) {\bf n}
\eeq
This suggests the definition of a {\it third} complex vector
\beq
{\bf P} \ = \ ({\bf t} + i{\bf b})
\eeq
 This leads to
$ {\bf P}_s \ = \  (K  -i  \tau) {\bf n}$.
Thus  a  {\it third} complex function
\beq
\chi(s,u) \ = \  (K - i \tau)
\eeq
  appears  in this case.
Using the  above definitions of {\bf P} and $\chi$
  given above in  the basic equations  (2.1) and (2.2),
  and proceeding the same way as in the previous formulations,
 we get
\beq
{\bf P}_s \ = \ \chi {\bf n}~~~;~~~
{\bf n}_s \ = - \frac{1}{2} (\chi^* {\bf P} + \chi {\bf P}^*),
\eeq
\beq
{\bf P}_u = iR_3 {\bf P}  - \gamma_3 {\bf n}
~~~;~~~
{\bf n}_u \ = \  \frac{1}{2} (\gamma^*_3 {\bf P} + \gamma_3 {\bf
P}^*)\
\eeq
where
\beq
\gamma_{3} \ = - (g- i \tau_{0})~~~;~~~
R_3 \ = \ -h = -\int (K \tau_{0}-\tau g) ds.
\eeq
Here we have used the last  equation in the set (2.4) to write the
last equality above, and the subscript $3$
corresponds  to  formulation (III).
>From Eqs.(4.13) and (4.14), setting ${\bf P}_{su} \ = \
{\bf P}_{us}$, and equating the coefficients of {\bf n} and {\bf P},
respectively, we get
\beq
\chi_u + \gamma_{3s} - i R_{3}~~ \chi \ = \ 0;
\eeq
\beq
R_{3s} \ = \ \frac{i}{2} (\gamma_{3}~~ \chi^* - \gamma_{3}^*~~ \chi).
\eeq
Once again, Eqs. (4.16) and (4.17) have the same form as Lamb's result (Eqs. (3.7) and (3.8)),
 as well as the AKNS compatibility conditions  (3.11) and (3.12). Here, the
identifications   are
\beq
q = \chi /2;~~~A = i R_{3};~~~B= -\gamma_{3}/2;~~~\zeta=0.
\eeq
 Thus using the same reasoning as  before, for suitable choices
 of $\gamma_3$  as functions of $\chi$ and its derivatives,
  Eq. (4.16) for $\chi$  can take the form of  known integrable
 equations.

 Now, as is clearly seen from Eqs. (3.6), (4.6) and (4.15), the
parameters $\gamma_n$, $n=1,2,3$ that arise in the
 three formulations, correspond to three
 complex functions involving different combinations  of  the curve
evolution parameters  $g,h$  and  $\tau_{0}$. Further, the
corresponding
complex functions $\psi$, $\Phi$ and $\chi$  that appear in the
formulations
 are also different
functions of $K$ and $\tau$. (See Eqs. (1.1), (4.3) and (4.12)
respectively.) Thus it is clear that
the three formulations   describe three  {\it distinct}
classes of curve motion. In the next section, we discuss
an application to  show this explicitly.

\section {An illustrative example:
The nonlinear\\ Schr\"{o}dinger equation}
\setcounter{equation}{0}

 Next, we consider the application of our results to the NLS.
 This example  is appropriate not only because the NLS was one of the
 first integrable equations to appear in the context of curve
motion \cite{hasi},
but also because it has applications \cite{ablo} in various fields such
 as vortex filament motion, optical solitons, magnetic chain dynamics etc.

\noindent {\small{\bf Formulation (I)}}: As Lamb has shown, the choice
\beq
\gamma_1  = -i \psi_{s},
\eeq
when used in Eqs. (3.8) leads to
\beq
R_{1}= \frac {1}{2}|\psi|^2.
\eeq
Substituting  this in Eq(3.7) yields the NLS:
\beq
i\psi_u+\psi_{ss}+ \frac{1}{2}|\psi|^2 \psi=0.
\eeq
 Using Eq. (1.1) in Eq. (5.1) and  equating the
 resulting expression  to $\gamma_1$ defined in (3.6), we get
 the following curve evolution parameters $g, h$ and $\tau_0$
 for this case:
\beq
g~~=~~-K\tau~~~;~~~
h~~=~~K_{s}~~~;~~~
\tau_{0}~=~(h_{s}+~~\tau~g)/K~~=~~
(K_{ss}/K)-\tau^{2}
\eeq

On the other hand, substituting the above expressions for $g$ and
$h$ in the first equation of (2.2) leads to $
{\bf t}_u = g{\bf n} + h{\bf b} = -K\tau {\bf n} + K_s {\bf b}$.
 By using Eqs. (2.1) and (2.2), it can be easily verified that the above
 equation  implies $ {\bf t}_u = {\bf t} \times {\bf t}_{ss}$.
 This is identical in form to the   Landau-Lifshitz (LL) equation
  ${\bf S}_u = {\bf S} \times {\bf S}_{ss}$ for the
 time evolution of the spin
vector ${\bf S}$ in a continuous Heisenberg ferromagnetic spin chain.
  In this formulation, {\bf t}, the tangent to the space
 curve  gets identified with {\bf S}. The  connection between the  NLS
 and the LL equation  is well known \cite{laks} by now.

\noindent {\small {\bf Formulation (II)}}: Here, the choice
$ \gamma_2  = -i \Phi_{s}$
when used in Eqs.(4.8) leads to
$R_{2}= \frac {1}{2}|\Phi|^2$.
Using this in Eq. (4.7) yields the following  NLS for $\Phi$:
\beq
i\Phi_u+\Phi_{ss}+ \frac{1}{2}|\Phi|^2 \Phi=0,
\eeq
which is identical in form to Eq. (5.3) obtained in
formulation (I). However,
 $\Phi = \tau~~ {\rm exp} [i \int K ds]$  here. Using this to find
the above expression for $\gamma_2$,  and equating it to the
expression for $\gamma_2$ defined in Eq. (4.6), we obtain
 the following curve evolution parameters:
\beq
h~~=~~-\tau_{s}~~~;~~~
\tau_{0}~~=~~-K~~\tau~~~;~~~
g~=~(K~~\tau_{0}-h_{s})/\tau~~=~~(\tau_{ss}/\tau) - ~~K^2.
\eeq
 Thus the above parameters $g$, $h$ and $\tau_{0}$
 which describe the curve motion in this formulation are quite
 different from those  of formulation (I) (See Eqs. (5.4)).
In addition, using the above expressions for $h$ and $\tau_{0}$ in the
 third  equation of Eqs. (2.2),  a short calculation shows that
for this curve, ${\bf b}_u =- {\bf b} \times {\bf b}_{ss}$.
 Thus, in contrast to formulation (I), it is the {\it binormal}
 of the moving curve that satisfies the LL equation
in this case.

\noindent {\small {\bf Formulation (III):}} Proceeding as in the previous
 two cases, here the choice is
$\gamma_3  = -i \chi_{s}$, implying
 $ R_{3}= \frac {1}{2}|\chi|^2$.  This  leads  to
\beq
i\chi_u+\chi_{ss}+ \frac{1}{2}|\chi|^2 \chi=0,
\eeq
where $\chi=(K -i \tau)$. (See Eq. (4.12)).
The curve evolution in this case is described by the three
parameters
\beq
g~~=~~\tau_s~~~;~~~
h~~=~~(1/2)(K^{2}~~+~~\tau^{2})~~~;~~~
\tau_{0}~~=~~-K_{s}.
\eeq

Thus by direct comparison of Eqs. (5.8), (5.4) and (5.6), we see that  the above expressions
for $g$, $h$ and  $\tau_{0}$,
 which describe the geometry of curve evolution in this case,
 are quite different
from  those obtained in formulations (I) and (II).
 Further, using  the parameters given in Eq. (5.8), it is possible
to show that the {\it  normal} to this moving curve satisfies the
LL equation in this case, i.e.,
  ${\bf n}_u = -{\bf n} \times {\bf n}_{ss}$. We paranthetically
remark that the correspondence  of
this equation to  the NLS equation  has been noted by Hirota
 in a different context \cite{hirota}.

 To demonstrate  that the three formulations
 lead to three
distinct space curve evolutions,  each  with a  different
 set $K$, $\tau$, $g$, $h$ and $\tau_{0}$,  we  consider
  the  simple  example  of a  one-soliton solution $q$ of the NLS
equation:

\beq
q~~=~~a_0~{\rm sech}~\xi
~~{\rm exp}~i~V_{e}(s- V_{c}u)/2.
\eeq

Here $V_e$  and $V_c$ denote, respectively, the envelope velocity
 and carrier velocity of the soliton. The
amplitude $a_{0}=[V_{e}(V_{e}-2V_{c}
)]^\frac{1}{2}$  and $\xi= (s-V_{e}u)(a_{0}/2)$.
 Also, $V_{e}(V_{e}-2V_{c})\ge 0$.
  Note that any two of the three parameters $a_0$,
 $V_e$ and $V_c$ can be
taken to be independent. We have
 taken the case $V_{c}=0$ for illustration, and
obtained the  expressions
 for $K$ and $\tau$ for this solution, in the three
 formulations, by using $\psi$, $\Phi$ and $\chi$ defined
 in Eqs. (1.1), (4.3) and (4.12) respectively.
 These have been entered in  Table 1. Using
 these, the corresponding  time-evolution parameters
 $g,h$ and $\tau_0$
 for the three formulations  are computed from
 Eqs. (5.4), (5.6) and (5.8) respectively.
 These are given in Table 2.

\noindent{\small\bf New geometries associated with a soliton solution of
NLS:}  We now present the geometrical consequences  that arise
 from the  three formulations, at the curve level.
Let  the moving curve be described by the vector $ {\bf r}(s,u)$. Thus the
tangent ${\bf t}$  appearing in the Frenet-Serret  equations
(2.1) is defined as ${\bf t}={\bf r}_{s}(s,u)$.
 According to the fundamental theorem of
curves  \cite{eisen},  smooth
functions $K (>0)$ and $\tau$ define a curve ${\bf r}(s,u)$ uniquely,
 modulo orientation in space. Now, from the expressions for
$K$ and $\tau$ given in Table 1, we see that
 formulation (I)  yields  a moving
space curve with constant torsion $\tau$ but a space-time varying
curvature $K$. On the other hand, the
  moving curve obtained using formulation (II)
  has a constant $K$ but varying $\tau$. In formulation (III),
 both $K$ and $\tau$   are space-time dependent.
 Since the three moving curves have different $(K,\tau)$
 parameters,
 our results  explicitly  illustrate that they
are indeed {\it geometrically distinct}. Further, since they
 are all gauge equivalent to  the NLS, they correspond to
 integrable curve evolution.

  Now,  any  moving curve  arising from
 formulation (I) is  well known to be the solution of the
 following  "localized induction (LI) equation"  \cite{hasi}
 for a vortex filament: ${\bf r}_u~=~{\bf r}_s\times{\bf r}_{ss}$.
 This is because the LI equation
  can be easily shown to imply the  LL equation
 for the tangent ${\bf t}$ that appears in formulation (I).
 In contrast, in formulations (II) and (III),
the LL equation is satisfied by ${\bf b}$ and ${\bf n}$
respectively, and hence  the corresponding ${\bf r}(s,u)$
 does not obey
  the LI equation any more.
 Thus the curves arising from  the two latter formulations
 are  distinct from that associated with the LI equation.
  It is of course, not at all obvious whether
 analogs  of the LI equation (i.e., a  PDE  at  the  curve level)
can be written down for these two  formulations. However,
 irrespective of this, it
   turns out that  there is a procedure  to  construct
 the corresponding  moving curve  ${\bf r}(s,u)$ itself,
 for each of these formulations.
 The results of this  lengthy calculation
  will be presented elsewhere.

  Clearly, the dynamics of
these new curves which   get associated with the  LL equations for
 {\bf n} and {\bf b}  respectively, would  also be form-preserving
  curves, just as Hasimoto's vortex
filament  was. These may  find applications
 in fields other than fluid dynamics. For instance, a phenomenological
 modelling  of the motion of  an interface (regarded as a curve)
 could be one possible application.

\section{Conclusions}
\setcounter{equation}{0}

 In this Letter,  we have   found new connections between
 moving space curves and solitons using  two natural extensions
 of Lamb's formulation. Our  unified procedure demonstrates
  that a specific  evolution of an integrable equation  can
  actually  be associated with {\it three} different classes of
 space curve motion, the class discussed  by Lamb being one of these.
As described in Sec. 3, the
 corresponding  Lax pairs in the three cases can also be
 constructed. (See Eq. (3.14) and the discussion
 below it \cite{fn1}.)

 Application to the NLS  is  discussed. Our  analysis
 shows that the class of curves that
 are solutions of the well known "localized induction equation"  for
 a vortex filament is but {\it one} of the three classes
 of moving curves that
  can be associated with the NLS, i.e., the class
  connected to Lamb's formulation. There are (as
 we have shown in general) two more classes, and
in order to understand these new  geometric structures of
 the NLS  more  explicitly,
  we have determined the  various curve parameters for each of the three
  moving curves associated with a one-soliton solution (5.9)
 of the NLS  and displayed them in Tables 1 and 2.
  While the curve corresponding to Lamb's class has a
constant torsion but varying curvature, the  second curve  has a
 constant curvature, but varying torsion. For the third curve, neither of
these parameters is a constant. As this example shows, these  three
are clearly distinct curves. It would indeed be instructive to apply
our results to  other integrable equations as well,  and
 find the three associated geometric structures.

 We conclude with a  general remark on the
  possible connection between a singularity in the
 "velocities" (i.e., the  time evolution parameters)
  of the geometric dynamics and that of the associated NLPDE.
  We have shown
 that the three complex functions
  $ \psi  =  K  \exp { [i\int \tau ds] }$,
$\Phi  =  \tau  \exp {[i\int K  ds]}$ and $
\chi  =  (K - i \tau)= (K^{2}+\tau ^{2})^{\frac{1}{2}}
\exp{[i\int \tan^{-1} (\tau/K) ds]}$, satisfy the same NLPDE,
  and   that three different moving curves get associated with it.
  Therefore if a certain solution of
the NLPDE has a singularity at a
 point, then it implies that  as the curve evolves, a corresponding singularity
 will appear in  the curvature  $K$  of the  first  moving curve, in
the torsion $\tau$ of the second curve, and
in the quantity $(K^{2}+\tau ^{2})^\frac{1}{2}$ for the third.
 Since these "velocities" $g,h$ and $\tau_{0}$
  will be  certain functionals of $K , \tau$ and their derivatives for the three
  curves, the corresponding
 singularity will be reflected in these velocities  as well, and can be
 studied case by case.

\newpage
\vskip 1cm

\begin{table}
\begin{tabular}{|c||c|c|c| }
\hline
Formulation & Solution of NLS &~~ $K$~~ &~~$\tau~~$\\
\hline
I & $\psi$ & $ V_{e} \rm sech \xi$& $V_e/2$\\
\hline
II & $\Phi$ & $V_e/2$ & $V_{e} \rm sech \xi$\\
\hline
III & $\chi$& $V_{e} \rm sech\xi cos V_{e} s$&$-V_{e}
 \rm sech \xi sin
V_{e}s$\\
\hline
\end{tabular}
\vspace{0.5cm}
\caption { Example: The curvature $K$ and torsion
$\tau$ for the special soliton
 solution  (Eq. (5.9))
 of the nonlinear Schr\"{o}dinger equation (NLS)
 for $\psi$, $\Phi$ and $\chi$ in the
 respective  formulations (I), (II) and (III).
 This soliton has a vanishing
 carrier velocity and an envelope velocity $V_e$.}
\label{tab:TAB1}
\end{table}
\begin{table}
\begin{tabular}{|c||c|c|c| }
\hline
Formulation& ~~$g$~~ &~~ $\tau_0$~~ &~~$h~~$\\
\hline
I & $(-V_{e}^2/2) \rm sech \xi$ & $(-V_{e}^2/2)
\rm sech^2 \xi$& $(-V_e^2/2)\rm sech \xi \rm tanh \xi$\\
\hline
II& $(-V_e^2/2) \rm sech^2 \xi$& $(-V_e^2/2) \rm sech
\xi$&$ (V_{e}^2/2) \rm sech \xi \rm tanh \xi$\\
\hline
III& $(V_e^2/2) \rm sech \xi \times $&$(V_e^2/2)
\rm sech \xi \times $&\\
&$(\rm tanh \xi \rm sinV_{e}s -2\rm cosV_{e}s)$&
$(\rm tanh\xi \rm cosV_{e}s+ 2\rm sinV_es)$&$(V_e^/2)\rm sech^2\xi$\\
\hline
\end{tabular}
\vspace{0.5cm}
\caption { The coresponding time evolution parameters $g$,
$\tau_0$ and $h$  in the three formulations (Eqs. (5.4), (5.6) and (5.8))
for the example in Table 1.}
\label{tab:TAB1}
\end{table}


\newpage

\begin{thebibliography}{99}
\bibitem{lamb}G.L. Lamb, J. Math. Phys.   18 (1977) 1654 .

\bibitem{ablo} See, for instance, M.J. Ablowitz and H. Segur,
Solitons and the Inverse Scattering Transform,
 SIAM, Philadelphia, PA 1981.

\bibitem{hasi} H. Hasimoto, J. Fluid. Mech.  51 (1972)  477 .

\bibitem{klapper}See, for e.g.,  I. Klapper , M. Tabor, J. Phys. A
27 (1994) 4919 ;  K. Nakayama, H. Segur , M. Wadati,
Phys. Rev. Lett.  69 (1992)  260;  R.E. Goldstein,
 S. A. Langer, Phys. Rev. Lett.  75 (1995)  1094
and references therein.

\bibitem{eisen} L.P. Eisenhart,  A Treatise on the
Differential Geometry of Curves and Surfaces, Dover, New York,
1960.

\bibitem{radha1} Radha Balakrishnan, A. R. Bishop,  R. Dandoloff,
 Phys. Rev. B  47 (1993) 3108; Phys. Rev. Lett.
64 (1990)  2107 ; Radha Balakrishnan , R. Blumenfeld, J. Math.
Phys.  38  (1997) 5878 .

\bibitem{radha2}Radha Balakrishnan, A. R. Bishop , R. Dandoloff, Phys.
Rev. B  47(1993)  5438 .

\bibitem{laks} M. Lakshmanan, Phys. Lett. A  61 (1977)  53 .

\bibitem{fn1} We caution that the cases $\alpha = 1,2,3$ (see
below Eq. (3.10)) should not be confused with the
 formulations (I), (II) and (III).

\bibitem{hirota} R. Hirota, J. Phys. Soc. Jap. 51 (1982) 323.
Here, $\chi$ was used to illustrate that the LL equation (for ${\bf n}$)
and the NLS (for $\chi$) transform to the same bilinear form.
\end{thebibliography}
\end{document}